\documentclass[twocolumn,pra,aps,showpacs]{revtex4}
\usepackage{amsmath}
\usepackage{epsfig}

\begin{document}

\title{Collective Excitations of a Bose-Einstein Condensate in an Anharmonic Trap}
\author{Guan-Qiang Li$^{1,2}$, Li-Bin Fu$^{1}$, Ju-Kui Xue$^{2}$, Xu-Zong Chen$^{3}$%
, and Jie Liu$^{1*}$}
\affiliation{1. Institute of Applied Physics and Computational Mathematics, P.O. Box 8009
(28), 100088 Beijing, China\\
2. Physics and Electronics Engineering College, Northwest Normal University,
730070 Lanzhou, China\\
3. Key Laboratory for Quantum Information and Measurements, Ministry of
Education, School of Electronics Engineering and Computer Science, Peking
University, Beijing 100871, Peoples Republic of China}

\begin{abstract}
We investigate the collective excitations of an one-dimensional
Bose-Einstein condensate  with repulsive interaction between atoms
in a quadratic plus quartic trap. With using variational approaches,
the coupled equations of motions for the center-of-mass coordinate
of the condensate and its width are derived. Then, two low-energy
excitation modes are obtained analytically. The frequency shift
induced by the anharmonic distortion, and the collapse and revival
of the collective excitations originated from the nonlinear coupling
between the two modes, are discussed.
\end{abstract}
 \maketitle

\section{Introduction}

One of the most important characters of an interacting quantum
many-body system is its response to external perturbations, where
collective excitation modes represent a very effective tool for
exploring the role of interactions and testing theoretical schemes.
For this reason, measurements of the collective modes in the trapped
gases of alkali-metal atoms \cite{c1,c2,c3} were carried out soon
after the discovery of Bose-Einstein condensates (BECs). For the
dilute degenerate gases, the essential physics of the BECs ground
state is included in the Gross-Pitaevskii equation (GPE). The
nonlinearity, originated from the interatomic interaction, is
included in the equation through a mean field term proportional to
condensate density. The study about the collective excitations  of
the condensation has been investigated extensively by using various
theoretical methods \cite{c4,c5,c6,c7,c8}. The remarkable agreement
between measured frequencies and theoretical predictions is one of
most important achievements in the investigation of these new
systems. Meanwhile, because of the emergence  of nonlinearity, a lot
of interesting phenomena in the collective excitations of BECs, such
as frequency shift \cite{c9}, mode coupling
\cite{c9,c10,c11,huang1}, damping \cite{c12,c13}, collapse and
revival of oscillations \cite{c14,c15} and the onset of stochastic
motions for strong driving amplitude \cite{c16,c16add, c16addadd},
have been paid much attention.

Recently, the collective dynamics of an one-dimensional trapped
ultra-cold Bose gases has attracted considerable attention, since
experiments on trapped Bose gases at low temperature have pointed
out the occurrence of characteristics 1D feature \cite{c17,c18,c19}.
In previous works the studies about the collective excitations of
BECs in the magnetic traps are mainly limited to the harmonic case.
However, in practical situation of experiments, the trap usually is
not purely harmonic. With this concern, we study the collective
excitations of the BECs in an one-dimensional in a harmonic trap
with a quartic distortion. Our aim is to understand how the
distortion affects the collective excitations of BECs.

Our study is facilitated by  variational approaches. Using a
Gaussian trial function, the GPE is transformed into a set of second
order ordinary differential equations about  some parameters that
characterize the condensate wave function. Then we derived the
expressions of the two low-energy oscillation modes analytically and
the nonlinear coupling between the two modes is revealed. In
particular, we find that, a very small anharmonic distortion may
cause a significant frequency shift of the excitation modes when the
atomic  interaction is strong. Finally, we demonstrate that the
anharmonic distortion may give rise to the collapse and revival of
the collective excitations.

The paper is organized as follows. In Sec.II we derive the governing
equations for the center-of-mass coordinate of the condensate and its width.
In Sec.III we discuss the collective modes and the frequency shift caused by
the anharmonic distortion. In Sec.IV, we demonstrate the collapse and
revival of the collective excitations in anharmonic potential. Final section
is our conclusion.

\section{ Variational approach and Governing equations}

We consider dilute degenerate bosons  confined in a cigar shaped
trap and assume that  the system is far from the Tonks-Girardeau
regime \cite{tg}. Then, the BECs can be well described by the
dimensionless 1D GPE,
\begin{equation}
i\frac{\partial \psi (x,t)}{\partial t}=[-\frac{1}{2}\frac{\partial ^{2}}{%
\partial x^{2}}+V(x)+g|\psi (x,t)|^{2}]\psi (x,t),  \label{nlse}
\end{equation}
where the coordinate $x$ is measured in unit of
$\sqrt{\hbar/m\omega_x}$ and time is in unit of $1/\omega_x$.
$\omega_x$ is the $x$ component frequency of the harmonic trap.
 $\psi (x,t)$ is the macroscopic wave function of the
condensate normalized so that $\int |\psi (x)|^{2}dx=1$; $g=4N\pi
\alpha _{1d}a_{s}/\sqrt{\hbar/m\omega_x}$ characterizes the
interatomic interaction and is defined in terms of the s-wave
scattering length $a_{s}$ (below we shall be concerned with
repulsive BECs for which $a_{s}>0$); $\alpha _{1d}=\int |\varphi
(y,z)|^{4}dydz/\left( \int |\varphi (y,z)|^{2}dydz\right) ^{5/2}$ is
a coefficient which compensates for the loss of two dimensions
\cite{c24}. In the above expressions, $N$ is the total number of
atoms and $\varphi (y,z)$ is the ground wave function of the lateral
dimensions.

The trapping potential we consider takes form
\begin{equation}
V(x)=\frac{1}{2}(x^{2}+\lambda x^{4}).
\end{equation}
The quartic term in the potential denotes the anharmonicity of the
trap. In \cite{c23}, the authors created such a quartic confinement
with a blue-detuned Gaussian laser directed along the axial
direction. In  their case, the non-rotating condensate was cigar
shaped and the strength of the quartic admixture was $\lambda \simeq
10^{-3}$. Here, we regard $\lambda $ as a controllable parameter and
assume that the  anharmonicity  is weak, i.e., $|\lambda| \ll 1$.

The problem of solving Eq.(\ref{nlse}) can be restated as a variational
problem corresponding to the minimization of the action related to the
Lagrangian density \cite{c6}
\begin{equation}
\ell =\frac{i}{2}(\psi \frac{\partial \psi ^{\ast }}{\partial t}-\psi ^{\ast
}\frac{\partial \psi }{\partial t})+\frac{1}{2}|\nabla \psi |^{2}+V(x)|\psi
|^{2}+\frac{g}{2}|\psi |^{4},  \label{lagrangian}
\end{equation}
where the asterisk denotes a complex conjugate. In order to obtain
the dynamics of the condensate in the trapping potential we will
find the extremum of Eq.(\ref{lagrangian}) with a set of trial
functions. In our case, a natural choice of the trial function is a
Gaussian, i.e., we take
\begin{equation}
\psi (x,t)=\eta (t)e^{-\frac{[x-\chi (t)]^{2}}{2w(t)^{2}}+ix\alpha
(t)+ix^{2}\beta (t)}.  \label{gaussian}
\end{equation}
At a given time $t$, this function defines a Gaussian distribution centered
at the position $\chi $ with width $w$. The other variational parameters $%
\eta $, $\alpha $ and $\beta $ are all real variables. Inserting
(\ref {gaussian}) into (\ref{lagrangian}) we can  calculate an grand
 Lagrangian by integrating  the Lagrangian density over whole
coordinate space, $L=\int_{-\infty }^{+\infty }\ell dx$. Then, from
the Lagrange equations, we obtain the evolution equations for all
variational parameters.

The dynamical equations for the center-of-mass and width of the condensate
is derived as follows
\begin{equation}
\ddot{\chi}+\chi +2\lambda \chi ^{3}+3\lambda \chi w^{2}=0,  \label{center}
\end{equation}
\begin{equation}
\ddot{w}+w+3\lambda w^{3}+6\lambda \chi ^{2}w=\frac{1}{w^{3}}+\frac{p}{w^{2}}%
,  \label{width}
\end{equation}
where the effective interaction $p\equiv \frac{g}{\sqrt{2\pi }}$, which
comes from the nonlinear interaction between the particles.

The other variational parameters can be obtained from the center
coordinate and the width through the equations,
\begin{equation}
\sqrt{\pi }|\eta (t)|^{2}w(t)=1,  \label{numconservation}
\end{equation}
\begin{equation}
\beta =\frac{\dot{w}}{2w},\hspace{0.5cm}\alpha =\dot{\chi}-\chi \frac{\dot{w}%
}{w}.  \label{other}
\end{equation}
The first one corresponds to the normalized condition of the wave function, $%
\int |\psi (x,t)|^{2}dx=1$. Therefore, once we know the behavior of
the center and width of the condensate, we can calculate the
evolution of the rest of the parameters, and then completely
characterize the dynamics of the condensate.

Comparing the above equations with that of a pure harmonic
potential, we  find the emergence of two new terms in
Eq.(\ref{center}) and (\ref {width}), i.e., the third and the fourth
terms in the left-side. Obviously, the third term is directly  from
the distortion of potential. Whereas, the fourth one represents  the
response of coherent wave to the distortion of potential manifesting
a  coupling between the  motions of center and width.

\section{Collective modes and frequency shift due to the anharmonic
distortion}

When we consider the contribution from the quartic distortion, i.e., $%
\lambda\neq0$ in the potential $V(x)$, the nonlinear coupling
between the  oscillations of the center and width  emerges. The
equilibrium points of Eq.(\ref{center}) and Eq.(\ref{width})
correspond to the stable or unstable stationary states of the
condensate. They satisfy following equations,
\begin{equation}  \label{center2}
\chi_{0}+2\lambda \chi_{0}^{3}+3\lambda \chi_{0} w_{0}^{2}=0,
\end{equation}
\begin{equation}  \label{width2}
(1+6\lambda \chi_{0}^{2})w_{0}+3\lambda w_{0}^{3}=\frac{1}{w_{0}^{3}}+\frac{p%
}{w_{0}^{2}}.
\end{equation}
There is only one stable equilibrium point for $\lambda>0$, that is,
\begin{equation}  \label{center3}
\chi_{0}=0,
\end{equation}
\begin{equation}  \label{width3}
w_{0}+3\lambda w_{0}^{3}=\frac{1}{w_{0}^{3}}+\frac{p}{w_{0}^{2}}.
\end{equation}
For $\lambda<0$ there are several equilibrium points, one of them is
stable and others are  unstable. The stable equilibrium point also
satisfies Eq.(\ref{center3}) and Eq.(\ref{width3}).

Expanding Eq.(\ref{center}) and Eq.(\ref{width}) around the
equilibrium points defined by (\ref{center3}) and (\ref{width3}) and
making a routine diagonalizing  process, we can obtain following
frequencies for  low-energy excitation modes:
\begin{equation}
\omega _{1,2}=\left[ 1+\left( 1\mp \frac{1}{2}\right) 6\lambda
w_{0}^{2}+\left( 1\mp 1\right) \left( \frac{3}{2w_{0}^{4}}+\frac{p}{w_{0}^{3}%
}\right) \right] ^{\frac{1}{2}}.  \label{freq}
\end{equation}
which are related to the coupled variation of center and width of
the condensate \cite{c4,c5,c6,c7,c8}. When $\lambda =0,$ $\omega
_{1}$ corresponds to the dipole oscillation $(m=1)$ characterizing
the motion of the center-of-mass, and $\omega _{2}$ is the frequency
of the variation of the condensate width, it is just the low-lying
collective mode $(m=0)$.

When the potential is not perfectly harmonic, i.e., $\lambda \neq
0,$ the
contribution from the quartic term will give rise to a  shift on the  frequencies. When $%
\lambda >0$ the frequency will be blue shifted and when $\lambda <0$
the frequency will be red shifted. This is true for both single
particle and BECs. However, it is interesting that for BECs, the
frequency shift is enhanced dramatically by the atomic interaction.
The frequencies of two low-energy excitations as the functions of
$p$ for above parameters are plotted in Fig.1.

\begin{figure}[!tbh]
\begin{center}
\rotatebox{0}{\resizebox *{9.0cm}{7.0cm} {\includegraphics {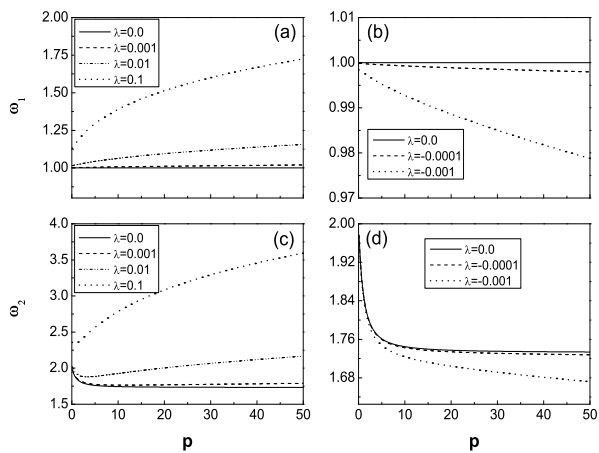}}}
\end{center}
\caption{Frequencies of two low-energy excitations as functions of effective
interaction $p$ for different $\protect\lambda$.}
\label{fig.1}
\end{figure}

From Eq.(\ref{freq}), we see that the contribution to the frequency
shift comes from the second term in the right-hang, i.e.,  $\sim
\lambda w_{0}^{2},$ which is due to the response of coherent wave to
the distortion of potential. On the other hand, the width\ $w_{0}$
of the wave function will be  broaden by the atom interaction. In
Fig. 2. we show this effetc by plotting the dependence of $w_{0}$ on
$\lambda $ for different interaction parameters.

\begin{figure}[!tbh]
\begin{center}
\rotatebox{0}{\resizebox *{9.0cm}{7.0cm} {\includegraphics {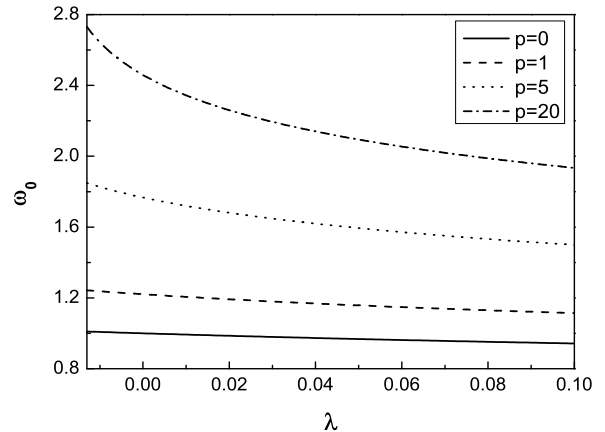}}}
\end{center}
\caption{The equilibrium width $w_{0}$ as functions of $\protect\lambda$ for
different effective interaction $p$.}
\label{fig.2}
\end{figure}

From the above discussion, we know, although the anharmonic
distortion is very small, the frequency shift maybe  large due to
magnification effects from the atom interaction. To demonstrate it,
in Fig.3 we plot the frequencies of dipole motion of BECs
wave-packet for different atom interactions and   anharmonic
parameters. It is clearly shown that,  the atom interaction will
give rise to $20\%$ or more shift of frequency when there is only
$2\%$ of the anharmonic distortion (see Fig.3, data at $p=20$). We
expect these phenomena can be observed in the future experiment.

\begin{figure}[!tbh]
\begin{center}
\rotatebox{0}{\resizebox *{9.0cm}{7.0cm} {\includegraphics {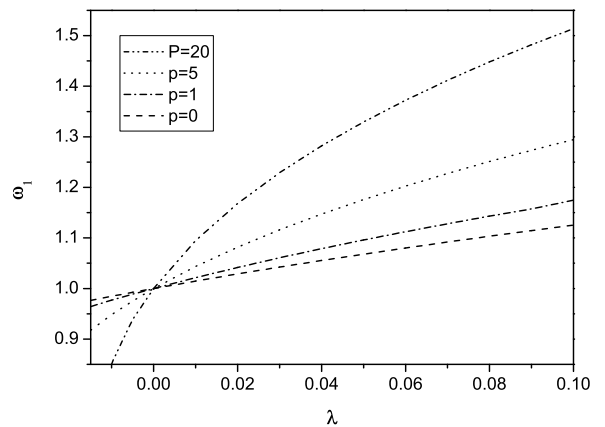}}}
\end{center}
\caption{Frequencies of mass center motions in anharmonic trap as functions
of $\protect\lambda$ for a classical single particle and BEC with different
atom interactions $p$.}
\label{fig.3}
\end{figure}

\section{Collapse and revival of the collective excitations}

\begin{figure}[!tbh]
\begin{center}
\rotatebox{-90}{\resizebox *{7.0cm}{9.0cm} {\includegraphics
{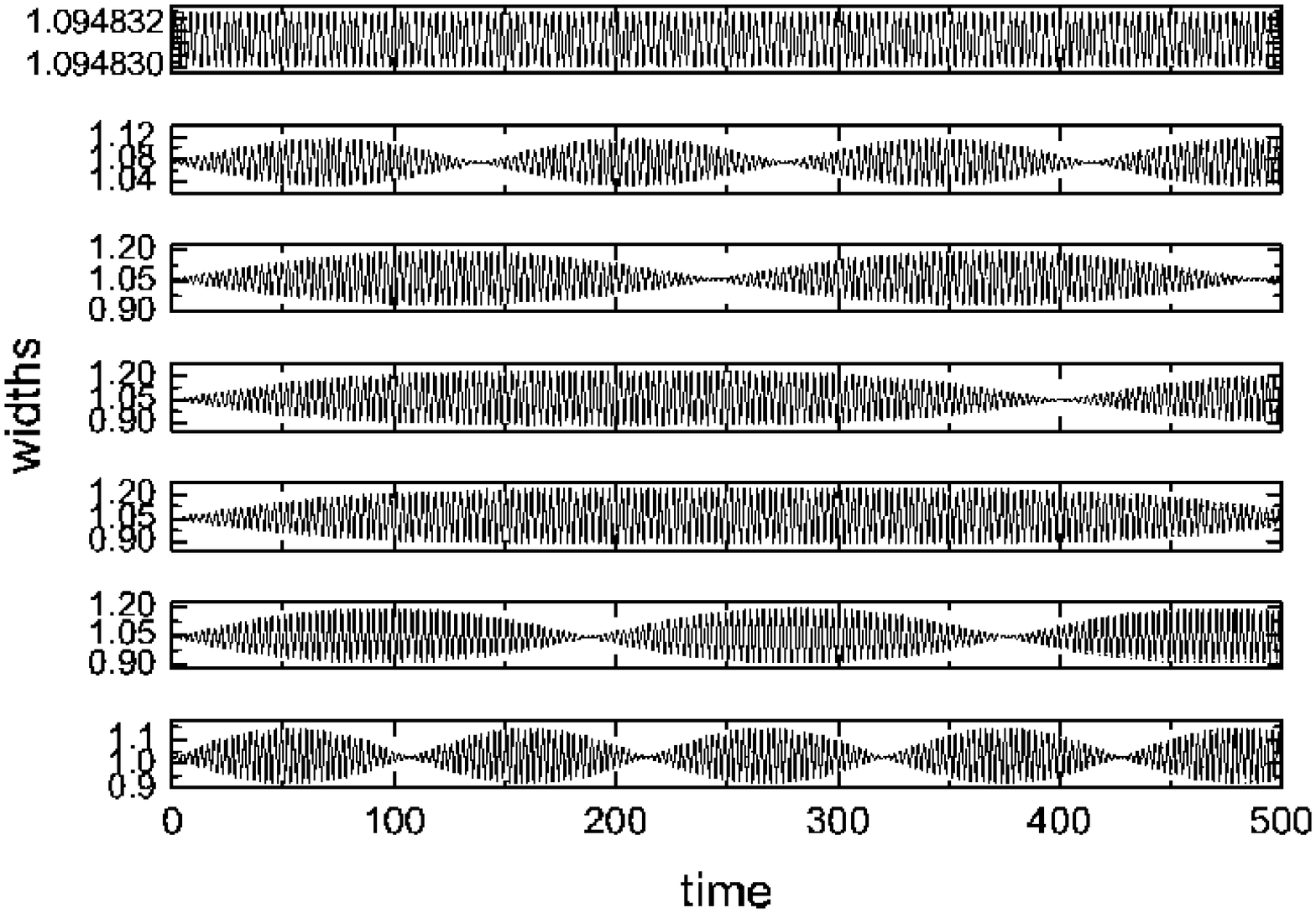}}}
\end{center}
\caption{Variations of the condensate width $w$ for the parameter $p=0.4$
and the different values of the anharmonic parameter $\protect\lambda$. From
up to down $\protect\lambda$=0, 0.02, 0.04, 0.05, 0.053, 0.06, 0.08.}
\label{fig.4}
\end{figure}

Another promising direction is to investigate the collapse and
revival of the collective excitations \cite{c14,c15}, which are
directly induced by the nonlinear coupling effect between two
oscillation modes. In the previous works, the nonlinear coupling
originates from the intrinsic interaction between particles in the
system. Here, we show a different mechanism for the nonlinear
coupling  that is due to the nontrivial anharmonic corrections to
the trap.

In our case, the qualitative results can be obtained with making
analysis on  Eqs.(\ref{center}) and (\ref{width}). The oscillation
of the center of the condensate couples with the motion of width
through the anharmonic parameter $\lambda $.  It is noted that the
coupled motion can be triggered by putting a small shift on the
center of the condensate from its equilibrium point. The consequent
motions of the width of the condensate excited by the shift are
illustrated in Fig. 4 for different anharmonicity parameters. In all
the above calculations,
initial shift is set as  $%
\triangle \chi =0.1$. Fig.4 clearly shows  the collapse and revival
of oscillation patterns for  the condensate width with respect to
time, induced by the anharmonicity.

Moreover,  the change of  collapse and revival are not monotonic
with increasing $\lambda $ and the revival period can be effectively
controlled by adjusting the anharmonicity parameter. In fact, by
linearizing Eqs.(\ref {center}) and (\ref{width}) around equilibrium
point, we  have $\ddot{\chi^\prime }+\omega _{1}^{2}\chi^\prime =0$,
and $\ddot{w^{\prime }}+\omega _{2}^{2}w^{\prime }=-6\lambda
w_{0}\chi ^{\prime }{}^{2}.$ Obviously, the motion of width behaves
like  a periodically driven oscillator. Since the frequency of the
'external force' is $2\omega _{1}$, and the intrinsic frequency of
the width oscillation is $\omega _{2}$, so the linear combination of
the  two frequencies gives the
frequency of the collapse and revival, that is, $|2\omega _{1}-\omega _{2}|$%
. Detailed analysis also suggest that the frequency of the collapse
and revival increases monotonically with the nonlinear parameter
$p$, but is not
monotonic with increasing $\lambda $. It will vanishes for some parameters, e.g. at $%
\lambda \approx 0.053$ for $p=0.4$. This is confirmed by our numerical simulations, as in Fig. 4,
around $%
\lambda \approx 0.053$ for $p=0.4,$ the period of collapse and
revival becomes much longer.

\section{Conclusion}

We have investigated collective excitations of a Bose-Einstein
condensate in an anharmonic trap with using variational approaches
and obtained the analytical expressions for the frequencies of the
low-energy excitations. It is shown that the two low-energy
excitation modes, corresponding to the variations of the center and
width of the condensate, couple with each other. The blue-shift and
red-shift on the excitation frequency caused by the anharmonic
distortion is revealed and found to be more dramatic in the case of
strong atomic interaction. Furthermore, the collapse and revival of
collective excitations in the anharmonic potential is discussed.  We
hope our theoretical results will stimulate the experiments in the
direction.

\section*{Acknowledgments}

This work was supported by National Natural Science Foundation of China
(No.10474008,10604009), Science and Technology fund of CAEP, the National
Fundamental Research Programme of China under Grant No. 2005CB3724503, the
National High Technology Research and Development Program of China (863
Program) international cooperation program under Grant No.2004AA1Z1220.

\end{document}